\documentclass[aps,prl,twocolumn,superscriptaddress,amsmath,amssymb,floatfix]{revtex4}
\usepackage{graphicx}
\usepackage{times}
\usepackage[colorlinks,linkcolor=red]{hyperref}
\usepackage{amsmath}
\usepackage{amssymb}
\usepackage{float}
\begin{document}

\title{Chaos in high-dimensional dynamical systems}
\author{Iaroslav Ispolatov}
\affiliation{Departamento de Fisica, Universidad de Santiago de Chile,
Santiago, Chile}
\email{jaros007@gmail.com}

\author{Michael Doebeli}
\affiliation{Department of Zoology and Department of Mathematics,
  University of British Columbia, Vancouver, BC V6T 1Z4 Canada}
\email{doebeli@zoology.ubc.ca}

\author{Sebastian Allende}
\affiliation{Departamento de Fisica, Universidad de Santiago de Chile,
Santiago, Chile}
\email{sebastian.allende@usach.com}

\author{Vaibhav Madhok}
\affiliation{Department of Zoology and Department of Mathematics, University of British Columbia,
  Vancouver, BC V6T 1Z4 Canada}
\email{vmadhok@gmail.com}

\def\T{\Theta}
\def\D{\Delta}
\def\d{\delta}
\def\r{\rho}
\def\p{\pi}
\def\a{\alpha}
\def\g{\gamma}
\def\ra{\rightarrow}
\def\s{\sigma}
\def\b{\beta}
\def\e{\epsilon}
\def\G{\Gamma}
\def\om{\omega}
\def\l{\lambda}
\def\f{\phi}
\def\w{\psi}
\def\m{\mu}
\def\t{\tau}
\def\c{\chi}
\newcommand{\sgn}{\operatorname{sgn}}

\begin{abstract}
For general dissipative 
dynamical systems we study what fraction of solutions exhibit
chaotic behavior depending on the dimensionality $d$ of the
phase space. We find that  a system of $d$ globally coupled ODE's with
quadratic and cubic non-linearities with random
coefficients and initial conditions, the probability
of a trajectory to be chaotic  increases universally from
$\sim 10^{-5} - 10^{-4}$ for $d=3$ to essentially one for $d\sim 50$. 
In the limit of large $d$, the invariant measure of the dynamical systems exhibits universal scaling that
depends on the degree of non-linearity but does not depend on the
choice of coefficients, and the largest Lyapunov exponent converges to
a universal scaling limit. Using statistical arguments, we provide analytical explanations for the
observed scaling and for the probability of chaos.
\end{abstract}

\maketitle

\vskip 0.5 cm
%begin{twocolumn}
In many standard texts, a transition from
classical concepts to statistical mechanics is justified by the prevalence of chaotic and ergodic behavior as more degrees
of freedom are considered.  However, quantitative details of
such transitions from integrability to chaos apparently remain
elusive.  In this paper we consider the fundamental question of the likelihood of chaos in 
 general dissipative dynamical systems in continuous time as a function of the dimension of phase space.  
 We note that existing results
 about the probability of chaos vs. dimensions in discrete maps
 \cite{kaneko1984, kaneko1989, sprott2006} as well as in a Hamiltonian system of locally-coupled oscillators
 \cite{mulansky2011} cannot be applied to dissipative dynamical
 systems in continuous time with an arbitrary degree of non-linearity and global coupling.  
In the following we show that the probability that the solution of a
generic $d$-dimensional system of ODEs with quadratic and cubic 
non-linearities (\ref{main1},\ref{main2},\ref{main3}) is chaotic universally increases from $\sim 10^{-4} -10^{-5}$ for $d=3$ to
essentially $1$ for large $d$. 
The results of our numerical investigations are then explained analytically, using a
combination of scaling and statistical methods. 
These results are an extension and generalization of an investigation of the prevalence of
chaos in the dynamics of high-dimensional phenotypes under frequency-dependent natural selection \cite{id14}. 
However, the
applicability and significance of our results is not limited to
biological evolution, and in principle extends to dynamical systems in statistical and nonlinear
physics, hydrodynamics, plasma physics, control theory, and social and
economic studies.

To investigate the statistics of 
trajectories, we numerically solve the following systems of equations
which contain second- and third-order nonlinear terms of a 
general form, 
\begin{align}
\label{main1}
\frac{dx_i}{dt}=\sum_{j=1}^d b_{ij} x_j + \sum_{j,k=1}^d
a_{ijk}x_j x_k - x_i^3, \; i=1,\ldots,d,
\end{align}
\begin{align}
\label{main2}
\frac{dx_i}{dt}=\sum_{j=1}^d b_{ij} x_j + \sum_{j,k=1}^d
a_{ijk}x_j x_k +\\
\nonumber
+ \sum_{j,k,l=1}^d c_{ijkl}x_j x_k x_l- x_i^5, \; i=1,\ldots,d,
\end{align}
\begin{align}
\label{main3}
\frac{dx_i}{dt}=\sum_{j=1}^d b_{ij} x_j + \sum_{j,k=1}^d
a_{ijk}x_j x_k +\\
\nonumber
+ \sum_{j,k,l=1}^d
c_{ijkl}x_j x_k x_l- x_i^3|x_i|, \; i=1,\ldots,d.
\end{align}
The coefficients $\{a\}, \{b\},$ and $\{c\}$ were randomly and independently drawn from 
Gaussian distributions with zero mean and unit variance. The last
highest-order terms, $-x_i^3,\; - x_i^5$, and $-x_i^3|x_i|$ were introduced to
ensure confinement of all trajectories to a finite volume of phase space,
thus excluding divergent scenarios. 
 In \cite{id14}, we integrated  system
(\ref{main1}) for each dimension $d$ using a 4th-order
Runge-Kutta method for 50
sets of the coefficients $b_{ij}$ and $a_{ijk}$, each with 4
sets of random initial conditions. This procedure was repeated for the
current work. For (\ref{main2}) and (\ref{main3}), the numerical
simulations are significantly more complex and computationally
extensive. We therefore integrated systems (\ref{main2}) and
(\ref{main3}) using a 5th-order Runge-Kutta adaptive step method for 50 - 100
sets of the coefficients $b_{ij}$, $a_{ijk}$ and $c_{ijkl}$ each with
only a single set of random initial conditions.
For each trajectory we determined the Largest Lyapunov Exponent (LLE)
by perturbing the trajectory by a small magnitude $\d x_0$ in a random
direction, integrating both trajectories in parallel for time $\tau$, measuring
the distance between trajectories $\d x_{\tau}$, rescaling the
separation between trajectories back to $\d x_0$, and continuing this
for the course of the simulation. The LLE was calculated as 
\begin{align}
\label{ll}
\l = \frac {1}{\t} \ln \left(\frac {\| \d x_{\t} \|}{\| \d x_{0} \|}\right),
\end{align}
and subsequently averaged over the trajectory.  The time of integration was chosen such that the
average LLE saturated to a constant value and it was usually not less
than $\sim 10^4/d^{\beta}$ with $\beta=2,\;3,\;9/2$ for
Eqs.(\ref{main1},\ref{main2},\ref{main3}). We explain this scaling below.  In cases when a trajectory
converged to a stable fixed point and the LLE was persistently negative,
the integration was stopped.  
By visually inspecting remaining trajectories we derived a fairly
robust criterium, observing that trajectories with $\l \sim d^{\beta}$
(with the proportionality coefficient being of order of 0.1) are
chaotic, while the trajectories with $\l \sim 1$ are
``quasiperiodic'', i.e. converging to a limit
cycle. Rather infrequent intermediate cases where inspected and
classified individually. For Eq.~(\ref{main1}) where simulations were
less computationally-expensive, we were able to implement more refined
method of estimating $l$, first allowing considerable time for 
the system to settle on the attractor and only then starting averaging
$\l$. In this case we concluded that trajectories with average LLE
$\l\geq 0.1$ can be considered chaotic, while trajectories with the
 $|\l| \leq 0.1$ are quasiperiodic. The
precise distinction between the quasiperiodic and
chaotic trajectories is unimportant to the main conclusion of our
paper as the fraction of quasiperiodic trajectories never exceeds 25\%
and vanishes in higher dimensions.
\begin{figure*}
%[t]\resizebox{18.1cm}{!}
\includegraphics[width=.9\textwidth]{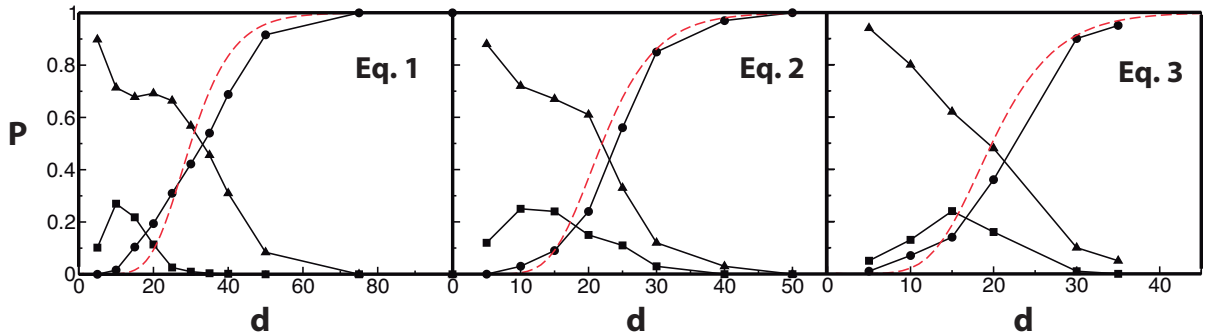}
\caption{Numerically measured probability of different types of dynamics as a function of
  dimension $d$ of the phase space for Eq.~(\ref{main1}) (left panel),
Eq.~(\ref{main2}) (central panel), Eq.~(\ref{main3}) (right panel):
$\bullet$ - chaotic trajectories, $\blacksquare$ -
limit cycles, $\blacktriangle$ - stable fixed points. For each case,
the theoretical estimate for the probability of chaotic trajectories (see main text)
is shown by a dashed line (red color online).}
\label{f1}
\end{figure*}

\begin{figure*}
%[t]\resizebox{18.1cm}{!}
\includegraphics[width=.9\textwidth]{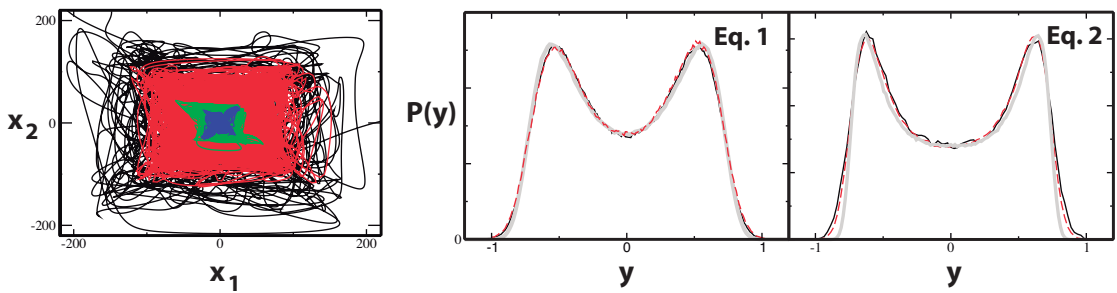}
\caption{\label{f2} Scaling of the size of chaotic
  trajectories (color online). Left panel: Examples of $x_1, x_2$
projections of   trajectories for the dynamics described by (\ref{main3})
for $d = 10$ (blue), $d = 15$ (green), $d = 30$ (red), and
$d=45$(black), illustrating the scaling $x_{i} \sim
d^{3/2}$. Central panel: The probability density for the scaled
coordinate $P(y)$ vs. $y=x/d^{\a}$, $\a=1$ of the solution of
Eq.~(\ref{main1}) for $d=150$ (solid black line), $d=100$ (dashed red
line), and the histogram of the solution of (\ref{stoch}) (thick grey line).
Right panel: The probability density for the scaled
coordinate $P(y)$ vs. $y=x/d^{\a}$, $\a=3/4$ of the solution of
Eq.~(\ref{main2}) for $d=65$ (solid black line), $d=50$ (dashed red
line), and the histogram of the solution of (\ref{stoch}) (thick grey line).
}\end{figure*}

\begin{figure}
\includegraphics[width=.35\textwidth]{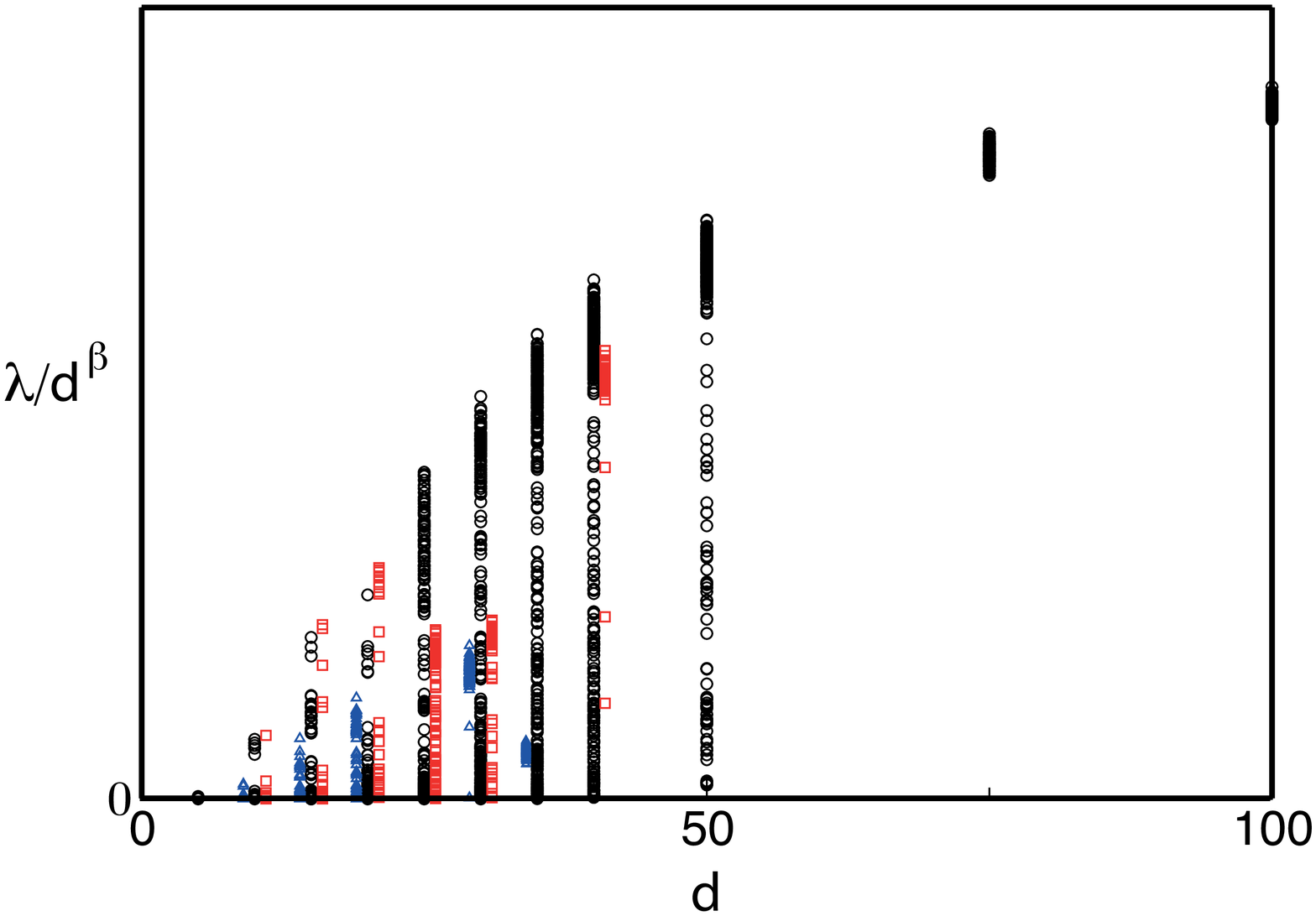}
\caption {\label{f3} 
 The scaled LLE  $\l/d^{\b}$  as a function of the dimension $d$ of
 phase space for (\ref{main1}); $\b=2$, black circle;  (\ref{main2}),
 $\b=3$, square, shifted to the right, red online; and (\ref{main3}),
 triangle, shifted to the left, blue online, $\b=9/2$. For large $d$,
 the LLE for  (\ref{main1}) extrapolates to $\l/d^{\b} \rightarrow \l^* \approx  0.235$ (see main text).
}\end{figure}

Our main result is that for all considered types of non-linearity the probability of chaos increases
with the dimension of the phase space, Fig.~\ref{f1}. In particular, the numerical simulations for 
(\ref{main1},\ref{main2},\ref{main3}) suggest that, essentially all trajectories become chaotic for $d\gtrsim 50$.
Our simulations also indicate that already for intermediate dimensions
$d\gtrsim15$, the
majority of chaotic trajectories essentially fill out the available
phase space, i.e., become ergodic (Fig.~\ref{f2}, left panel). In such a regime the
probability density $P(x_i)$ for each coordinate of the chaotic attractor
approaches a universal scaling form that depends neither on the
choice of coefficients  $\{a\},\{b\},\{c\}$ nor on the dimension $d$, Fig.~\ref{f2}. Furthermore, the LLEs also exhibits apparent scaling behavior, Fig.~\ref{f3}.

Below we explain  the
scaling and statistical properties of the large-$d$ limit of
(\ref{main1},\ref{main2},\ref{main3})
First,  consider the scaling of the spatial 
coordinates $x_i\sim d^{\a}$ and the LLEs $\l \sim d^{\b}$ illustrated in Figs.~\ref{f2} and \ref{f3}.
Consider the general case of a dynamical system similar to
(\ref{main1},\ref{main2},\ref{main3}) with the $n$th-order highest nonlinear term
and the $|x_i|^m \sgn(x_i), m>n$ diagonal confining term,
\begin{align}
\label{gen}
\frac{dx_i}{dt}=\sum_{k=1}^n\sum_{j_1,\ldots,j_k=1}^d g_{i,
  j_1,\ldots,j_k}^{(k)} x_{j_1}\ldots x_{j_k} - |x_i|^m \sgn(x_i).
\end {align}
Since the coefficients $g_{i, j_1,\ldots,j_k}^{(k)}$  in (\ref{gen}) are drawn randomly, it is reasonable to assume that each coordinate has a similar
scale, $x_i \sim x$ and  (\ref{gen}) becomes
\begin{align}
\label{scale0}
\frac{dx}{dt} \sim \sum_{k=1}^n x^k \sum_{j_1,\ldots,j_k=1}^d g_{i,
  j_1,\ldots,j_k}^{(k)}  - |x|^m \sgn(x).
\end{align}
Here the $g_{i, j_1,\ldots,j_k}^{(k)}$ are identically distributed random
terms with zero mean and unit variance, and a typical value of the sum of $N$ such terms is the standard deviation $\sqrt{N}$, which yields the following scaling relation:
\begin{align}
 \label{scale}
\frac{dx}{dt} \sim \sum_{k=1}^n x^k d^{k/2}  - |x|^m \sgn(x).
\end{align}
Introducing new variables,
\begin{align}
\label{scale_u}
y&= \frac{x}{d^{\a}},\;\a=\frac{n}{2(m-n)}  \\
\nonumber
\theta &= t d^{\b},\; \b=\frac{n(m-1)}{2(m-n)}
\end{align}
we convert (\ref{scale}) into
\begin{align}
\label{scale_eq}
\frac{dy}{d\theta} \sim \sum_{k=1}^n y^k d^{m(k-n)/[2(m-n)]}  - |y|^m\sgn(y).
\end{align}
On the right-hand side of (\ref{scale_eq}),  the highest-order $k=n$ term
and the $|y|\sgn(y)$ term do not depend on $d$ while the lower-order
terms with $k<n$ 
vanish in the limit of $d \gg 1$. 
The transformation (\ref{scale_u}) explains the observed scaling of
the size of chaotic attractors and the LLEs (whose dimension is the
inverse of time) shown in Fig.~\ref{f2} and \ref{f3}.  A more detailed
example of the above derivation for Eq.~(\ref{main1}) is given in \cite{id14}.
%%%%%%%%%%%%%%%%%%%%%%%%%%%%%%%%%%%%%%%%%%%%%%%%%%%%%%%%
To explain the shape of the universal probability density  $P(y)$
shown in Fig.~\ref{f2}, we ignore the irrelevant low-order terms  and replace the
leading nonlinear $n$th-order term (quadratic in (\ref{main1}) and cubic in
(\ref{main2},\ref{main3})) by a stochastic function $f(\theta)$. This is done observing that for large $d$, the majority of the
terms comprising the $\sum_{j_1,\ldots,j_n=1}^d g_{i,
  j_1,\ldots,j_n}^{(n)}$
 do not contain $x_i$ and can be approximated as
independent random variables. Since  $\langle
g^2 \rangle=1$ by definition, it follows from the Central Limit
theorem that
this sum is a Gaussian random variable with variance $D=d^n \langle
x^2 \rangle^n$. 
This leads to the following approximation of (\ref{scale0}) in the rescaled variables $y$ and $\theta$ of (\ref{scale_u}):
\begin{align}
\label{stoch}
\frac {dy} {d\theta} = f(\theta) - y^m,
\end{align} 
where $f(\theta)$ is  Gaussian process with dispersion $D$. To calculate the invariant measure of this process we approximate $f(t)$ by a jump
process which takes constant Gaussian-distributed values $f_i$ during
time intervals drawn from a uniform distribution with an average period
$\tau$. We solve to the scaling equation (\ref{stoch})
self-consistently, computing $\langle y^2 \rangle^n$
from the histogram of the trajectory $y(\theta)$ produced via
(\ref{stoch}). 
Varying $\tau$, we
find the best fit to the observed $P(y)$, which is shown as dashed
lines in Fig.~\ref{f2}. 
Given the approximate nature of the temporal behaviour of
$f(\theta)$ the fit seems quite satisfactory and yields $\tau^{(1)}=3.85$
for (\ref{main1}), $\tau^{(2)}=5.64$ for (\ref{main2}), and $\tau^{(3)}=6.63$ for (\ref{main3})
 Note that the estimate for $\tau^{(1)}=3.85$ is in a qualitative
 agreement with the large-$d$ asymptotic value of the corresponding rescaled
 LLE  $1/\l^*\approx 4.26$, (see Fig.~\ref{f3}), which characterizes the
 typical correlation time of the system. 
 
Next we provide a statistical explanation for the probability
of chaos as a function of the  dimension $d$, as illustrated  in ~\ref{f1}.
Consider stationary points of the dynamical systems
(\ref{main1},\ref{main2},\ref{main3}). Since a
system of $d$ $m$th-order algebraic equations generally has $md$
solutions (sometime coinciding), the dynamical system   (\ref{main1}) has $md$
stationary points $x^*$.  For our derivation, we assume that the system is chaotic if all these stationary points are
unstable in at least one direction, i.e., if at each stationary
point  $x^*$ at least one
eigenvalue of the local Jacobian matrix $J(x^*)$ 
has a
positive real part.  We assume that for sufficiently high $d$, all Jacobian eigenvalues are
statistically independent. This assumption of weakening correlations
between dimensions as the number of dimensions increase is a rather strong approximation without which is seems impossible to derive analytical estimates, and which seems to result in reasonable results (see below).
Denoting  the probability that the real
part of an eigenvalue is negative by $P_{neg}$, the probability that at least one out of
$d$ eigenvalues of the Jacobian at a stationary point has a positive real part
is $1-P_{neg}^d$. Hence the probability of chaos is 
\begin{align}
\label{P_chaos}
P_{chaos}=(1-P_{neg}^d)^{md},
\end{align}
indicating that for any $P_{neg}=1-\e<1$, 
the system becomes predominantly chaotic for $d\gtrsim1/\e$. Specifically, consider the
example of system (\ref{main2}) with cubic non-linearities.  
If $x^*$ is a stationary point of (\ref{main2}), the elements of the Jacobian
matrix $J(x^*)=\left\{J_{ij}(x^*)\right \}_{i,j=1}^d$ consist of two terms, 
 \begin{align}
\label{jacobian}
J_{ij}(x^*)&=\sum_{k=1,l=1}^d (c_{ijkl} +c_{iljk}+c_{ilkj} )x_k^{*}x_l^{*} -5x_i^{*4}\delta_{ij}\\
\nonumber
&\equiv J_{ij}^{(1)}+J_{ij}^{(2)},
\end{align}
where $\left\{\d_{ij}\right\}$ is the identity matrix.
As above, we ignored low-order terms present in (\ref{gen}), because such terms are irrelevant for large $d$. 
We assume that  the distribution of $x^*_i$  is the same as
for the coordinates $x_i$ themselves and is given by the universal
invariant measure shown in Fig.~\ref{f2}. We also consider the two terms $J_{ij}^{(1)}$ and $J_{ij}^{(2)}$ as statistically independent. The first
term, $J_{ij}^{(1)}$, is a sum
of $3d^2 \gg 1$ random variables with zero mean and a finite variance. 
Taking into account that the dispersions of $c_{ijkl}$ are one, and $x_i$
and $\{c_{ijkl}\}$ are uncorrelated (this follows from the observed independence of
$P(x)$ and the choice of $\{c\}$)
the Central Limit theorem states that this sum is a Gaussian-distributed variable with zero
mean and dispersion $\sigma^2 = 3d^2 \langle x^2 \rangle^2$. It follows from 
``Girko's circular law'' \cite{girko1984} that eigenvalues of a random 
$d \times d$-matrix with Gaussian-distributed elements with zero mean and unit variance are uniformly
distributed on a disk in the complex plane with radius $\sqrt{d}$. Thus, the eigenvalues of
$J_{ij}^{(1)}$ are  uniformly distributed on a disk with radius
$\sigma\sqrt{d}$. The
probability for an eigenvalue of $J_{ij}^{(1)}$  to have real part $r\sigma\sqrt{d}$,  with $|r| \leq 1$, is then
proportional to the length of the chord intersecting the radius of the disk at the point
$r$,
\begin{align}
\label{eigen}
P_c(r)=\frac{2\sqrt{1-r^2}}{\pi}.
\end{align}
(The factor $2/\pi$ normalizes $P_c(r)$ to one.)
 The probability distribution of the second, diagonal,  term of the Jacobian, $J_{ij}^{(2)}=- 5
x_i^{*4} \d_{ij}$ is defined by the invariant measure  $P(y)$,
given by (\ref{stoch}) and shown in Fig.~\ref{f2}. 
%%%%%%%%%%%%%%%%%%%%%%%%%%%%%%%%%%%%%%%%%%%%%%%%%%%%%%
It follows from scaling (\ref{scale_u}) that both $J_{ij}^{(1)}$ and $J_{ij}^{(2)}$ contribute terms of order $d^3$
to the eigenvalues of the Jacobian. The contribution from
$J_{ij}^{(1)}$ may have a positive or a negative real part with equal probability $1/2$. The
contribution from $J_{ij}^{(2)}$ is always negative and has magnitude
$5y^4$ with probability $P(y)$. It follows that the probability that the sum of
the two contributions has negative real part is 
\begin{align}
\label{ps1}
P_{neg}=\frac{1}{2} + \int\limits_{-\infty}^{+\infty} P(y) dy \int\limits_0^{5y^4/\chi} P_c(r)dr.
\end{align}
where $\chi\equiv \sigma /d^{5/2}=\sqrt{3} \langle
  y^2 \rangle$.
Integration on $dr$ produces
\begin{align}
\label{ps2}
P_{neg}=\frac{1}{2} \left[ 1 + \int\limits_{|y|>(\chi/5)^{1/4}} P(y) dy \right] \\
\nonumber
+\int\limits_{|y|<(\chi/5)^{1/4}} 
\frac{\sin^{-1}(5y^4/\chi) + 5y^4/\chi
  \sqrt{1-(5y^4/\chi)^2}}{\pi}P(y) dy.
\end{align}
Using the numerical data for $P(y)$ shown in Fig.~\ref{f2} we calculate
$\chi\approx 0.446$
and perform numerical integration of $P(y)$ to obtain $P_{neg}^{(2)}\approx
0.794$. A similar analysis for Eqs.~(\ref{main1}) (\cite{id14}) and
(\ref{main3}) yields $P_{neg}^{(1)} \approx
0.849$ and $P_{neg}^{(3)} \approx
0.787$, respectively.
Substituting these values into Eq.~(\ref{P_chaos}) provides a reasonable fit for the
observed probability of chaos, as illustrated by the dashed lines in
~\ref{f1}. An increasing discrepancy for lower $d$ could be attributed to
the facts that the systems reach truly scaling regime for $d
\rightarrow \infty$ and the histograms in Fig.~\ref{f2} are measured for
rather high  $d\geq 45$.

To summarize, we have presented numerical evidence that the behaviour
of generic dissipative dynamical systems in continuous time
universally becomes chaotic and ergodic as the
dimension of the phase space becomes large ($d \sim 50$ in the three
cases we studied). We note that the quadratic and cubic non-linearities considered
here can be interpreted as the first few non-linear terms in the
expansion of more complex non-linear dynamical systems, possibly
extending the applicability of our results. 
We have also provided 
some analytical explanations for the observed ubiquity of chaos and
for the universality of the density distribution of chaotic
trajectories. The similarity of the three panels in Fig.~\ref{f1} and the
apparently general applicability of Eq.~(\ref{P_chaos}) suggest that the
observed transition to chaos is not limited to the three
cases considered here and instead universally occurs in all
high-dimensional nonlinear dissipative dynamical systems. One of the goals of this work was to illustrate the
transition to chaos and ergodicity in high-dimensional phase space, a
frequently used yet rarely precisely stated argument in the formal justification of statistical
mechanics. Nevertheless, to explain our results we 
use scaling and probabilistic arguments borrowed from statistical physics. Thus, our results are an attempt to use statistical physics
to establish a basic ``phase diagram''  of dynamical systems. 

{\bf Acknowledgement} I.I. was supported by FONDECYT 1110288. M.D. was supported by NSERC, Canada. S. A. acknowledges financial support from FONDECYT 11121214 and 1120356, Grant  ICM P10-061-F by FIC-MINECON, Financiamiento Basal para Centros Cient\'ificos y Tecnol\'{o}gicos de Excelencia FB 0807, and  Concurso Inserci\'{o}n en la Academia-Folio  791220017.

\end{document}